\documentclass[conference]{IEEEtran}
\usepackage[numbers]{natbib}
\usepackage{gensymb}

\ifCLASSINFOpdf
   \usepackage[pdftex]{graphicx}
  \graphicspath{{images/}{../jpeg/}}
  \DeclareGraphicsExtensions{.pdf,.jpeg,.png}
\else
\fi
\usepackage[cmex10]{amsmath}
\ifCLASSOPTIONcompsoc
  \usepackage[caption=false,font=normalsize,labelfont=sf,textfont=sf]{subfig}
\else
  \usepackage[caption=false,font=footnotesize]{subfig}
\fi
\usepackage[hyphens]{url}

\begin{document}
\bstctlcite{IEEEexample:BSTcontrol}

%
\title{The statistics of low frequency radio interference at the Murchison Radio-astronomy Observatory}

\author{
\IEEEauthorblockN{Marcin Sokolowski}
\IEEEauthorblockA{ICRAR/Curtin University\\
GPO Box U1987\\Perth, WA, 6845. Australia\\
ARC Centre of Excellence for\\
All-Sky Astrophysics (CAASTRO)\\
Email: marcin.sokolowski@curtin.edu.au}
\and
\IEEEauthorblockN{Randall B. Wayth}
\IEEEauthorblockA{ICRAR/Curtin University\\
GPO Box U1987\\Perth, WA, 6845. Australia\\
ARC Centre of Excellence for\\
All-Sky Astrophysics (CAASTRO)\\
\and
\IEEEauthorblockN{Morgan Lewis}
\IEEEauthorblockA{School of Electrical and\\Computer Engineering\\
University of Western Australia\\
Crawley, WA 6009, Australia}
}}


%

\maketitle

\begin{abstract}
We characterize the low frequency radio-frequency interference (RFI) environment at the  Murchison Radio-astronomy Observatory (MRO), the location selected for the low-frequency component of the Square Kilometre Array. Data were collected from the BIGHORNS instrument, located at the MRO, which records a contiguous bandwidth between 70 and 300\,MHz, between November 2014 to March 2015 inclusive. The data were processed to identify RFI, and we describe a series of statistics in both the time and frequency domain, including modeling of the RFI occupancy and signal power as a series of distribution functions, with the goal of aiding future scientists and operation staff in observation planning.
\end{abstract}


%
\IEEEpeerreviewmaketitle

\section{Introduction}

Radio-frequency interference (RFI) is an ongoing challenge for radio astronomy where the difference in received power between astronomical radio sources and human-generated sources can be many orders of magnitude.
RFI can affect radio astronomy observations in several ways; at its most extreme, strong RFI saturates receivers rendering all data unusable.
Moderate power RFI can contaminate some frequency ranges within an observation while leaving other frequencies usable.
Low power RFI may only be detectable in long integrations of data where it is hard to excise.
The key science programs of current and future radio telescopes, including the Square Kilometre Array (SKA), require thousands of hours of observing time \citep[e.g][]{2013PASA...30...31B,2015aska.confE...1K}, hence understanding the properties of the RFI environment at these telescopes will help telescope operations staff and astronomers plan observations and data reduction processes.

The Murchison Radio-astronomy Observatory (MRO) is located in a remote area of Western Australia some 600\,km from Perth, approximately 300\,km from the nearest city of Geraldton (population about 36000) and about 90\,km from Meekatharra (population about 800), which is the nearest of local towns (with populations below 1000).
Most radio telescopes have a mechanism for monitoring the RFI environment. At the MRO, several low frequency radio telescopes are operating, including the Murchison Widefield Array (MWA) \cite{2013PASA...30....7T}, a prototype instrument for the future SKA-Low telescope \cite{2015arXiv151001515S} and BIGHORNS \cite{2015PASA...32....4S}.
Although the MRO is in a designated radio quiet zone with significant regulatory protections, RFI from aircraft and satellite-based transmitters is detectable as well as occasional weak FM radio and television broadcasts from distant transmitters.
Previously, \cite{2015PASA...32....8O} used 10 nights of data taken from an MWA sky survey project \cite{2015PASA...32...25W} to infer RFI statistics of the sky as observed by the MWA between 73 and 230\,MHz.
The observation strategy of the survey, however, breaks the frequency range up into five sub-bands of approximately 31\,MHz, hence the duty cycle at any one frequency was only 20\%. In addition, the phased-array antennas of the MWA are steerable hence can enhance or exclude RFI depending on where it comes from (i.e. near the horizon vs near the zenith).

In this paper, we use data collected from the BIGHORNS system between 2014 November and 2015 March (inclusive) to further understand the properties of RFI at the MRO.
Our analysis uses both the calibrated total power and the data ``flags'' (i.e. data marked as being affected by RFI), which are generated by automated processing of data collected by the BIGHORNS system, to identify RFI and study their statistics. 
The statistics of RFI was studied in terms of occupancy calculated in all frequency channels in the analyzed band, which for a single channel we define as a number of measurements flagged as RFI divided by a total number of measurements.

\section{Data acquisition and RFI detection}
\label{sec:data_acq_analysis}
Data were recorded with the BIGHORNS system which we briefly review here and refer the reader to \cite{2015PASA...32....4S} for details.
BIGHORNS is a calibrated total power spectrometer, attached to a single antenna, that records the power detected from the sky with approximately 117.2\,kHz frequency resolution and 0.05\,s time resolution.
Although the BIGHORNS system was designed as a science experiment, it is very useful as a general purpose radio spectrum monitor.

During the period of observations, BIGHORNS used a bespoke conical log-spiral (CLS) antenna that is well matched to left-hand circularly polarized (LHCP) emission from the sky between 50 and 350\,MHz, as detailed in \cite{2015ApJ...813...18S}.
Filters in the BIGHORNS signal chain limit the useful upper frequency to approximately 300\,MHz and ripples in the antenna frequency response limit the lower frequency to approximately 70\,MHz.
Data were calibrated according to the standard procedure developed for the BIGHORNS instrument \cite{2015PASA...32....4S,2015ApJ...813...18S}.

\begin{figure}[]
\centering
\includegraphics[width=\columnwidth]{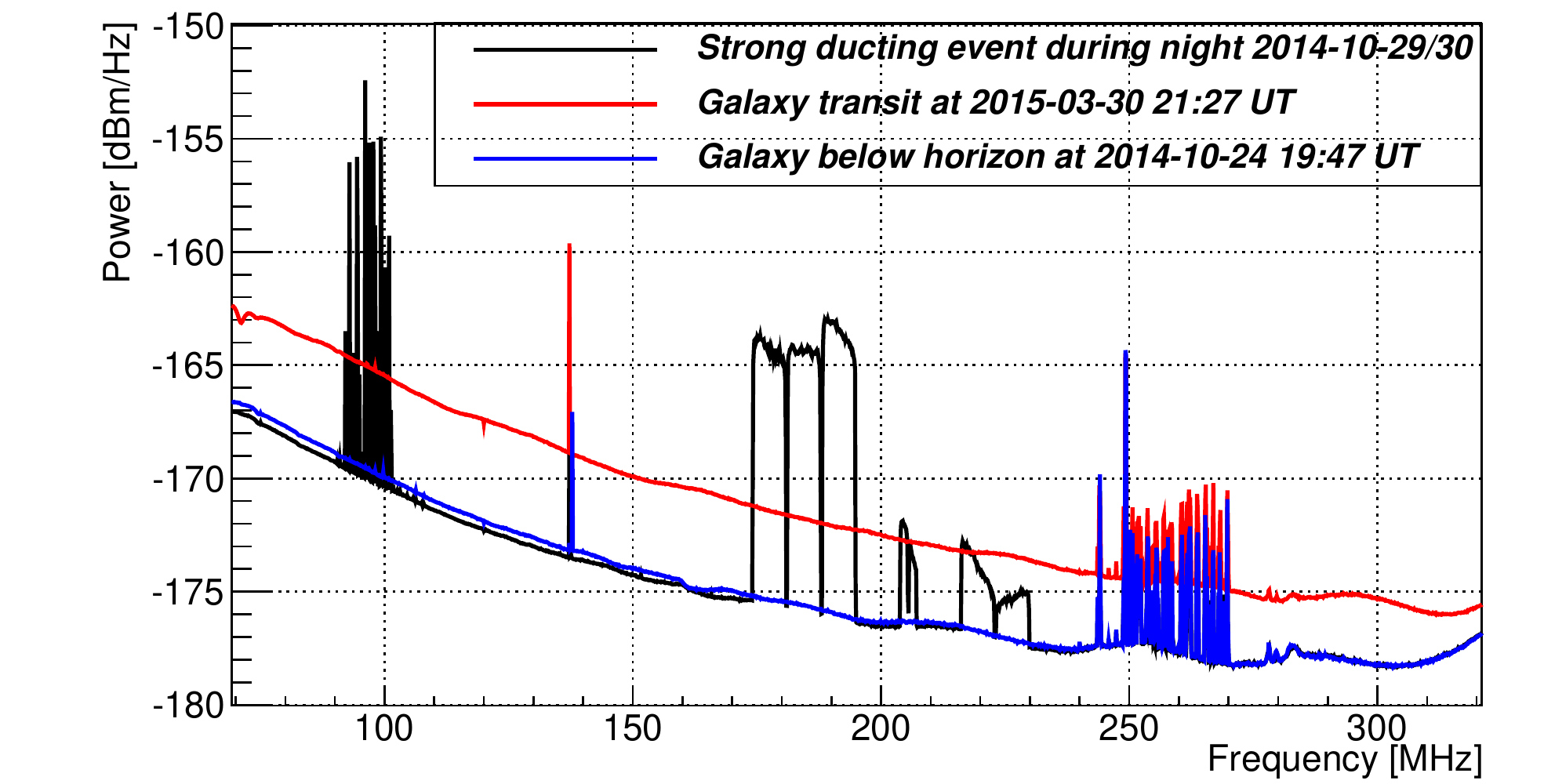}
\caption{Example nighttime spectra from the BIGHORNS system when the Galactic plane is above and below the horizon (red and blue lines). The black curve is a spectrum collected during an extreme RFI event, presumably caused by tropospheric ducting, with digital TV and FM radio signal more than 10\,dB above the typical Galactic noise level. Persistent satellite-based RFI is seen between approximately 242 and 272\,MHz. Intermittent strong RFI from Orbcomm satellites is seen around 137\,MHz.}
\label{fig:example_spectra}
\end{figure}

\begin{figure}[]
\centering
\includegraphics[width=\columnwidth]{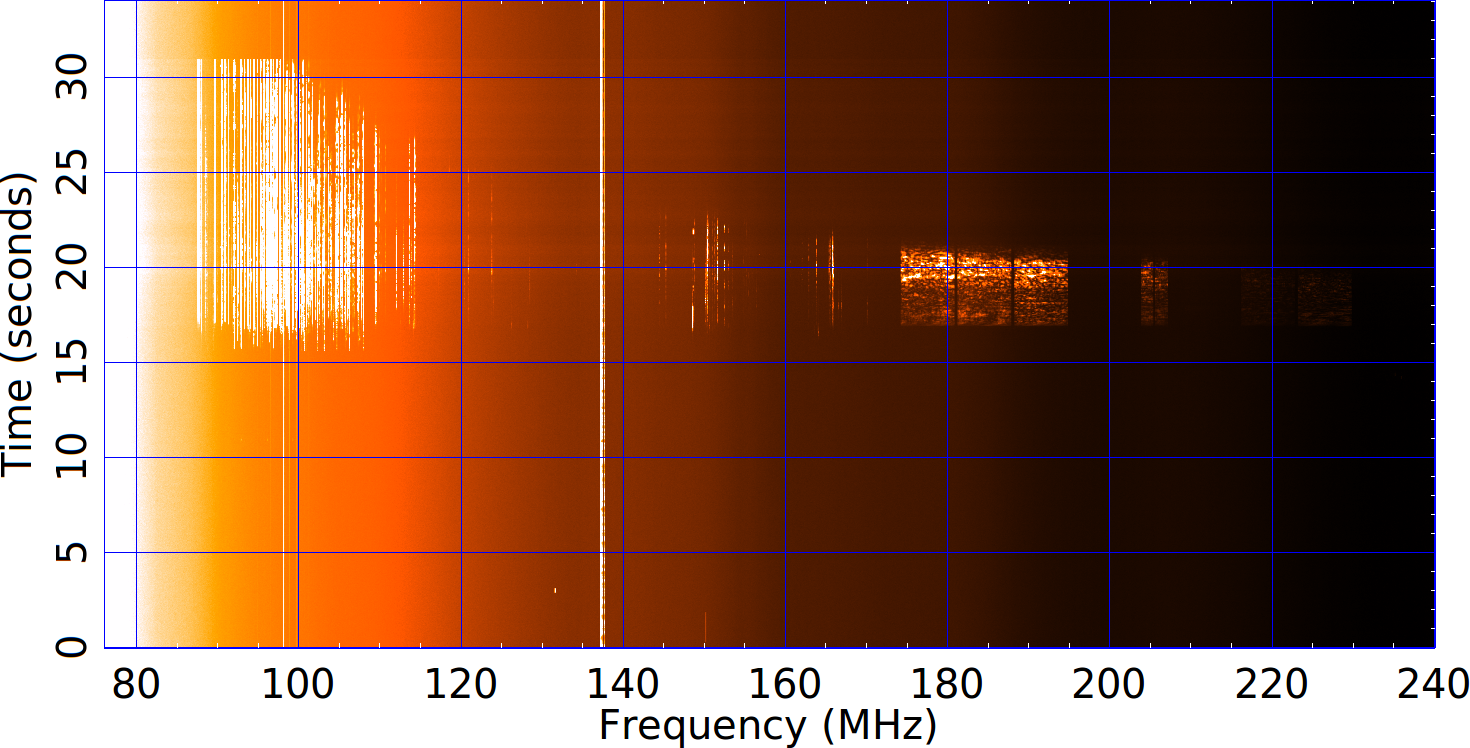}
\caption{Example of dynamic spectrum (frequency on horizontal axis, time on vertical axis and power indicated by colour) with an example of broadcast transimission (in DTV/DAB and FM bands) reflection of the aircraft or meteor (observation started at 2015-03-18 14:05:15 UT).}
\label{fig_reflection_example}
\end{figure}

\begin{figure}[]
\centering
\includegraphics[width=\columnwidth]{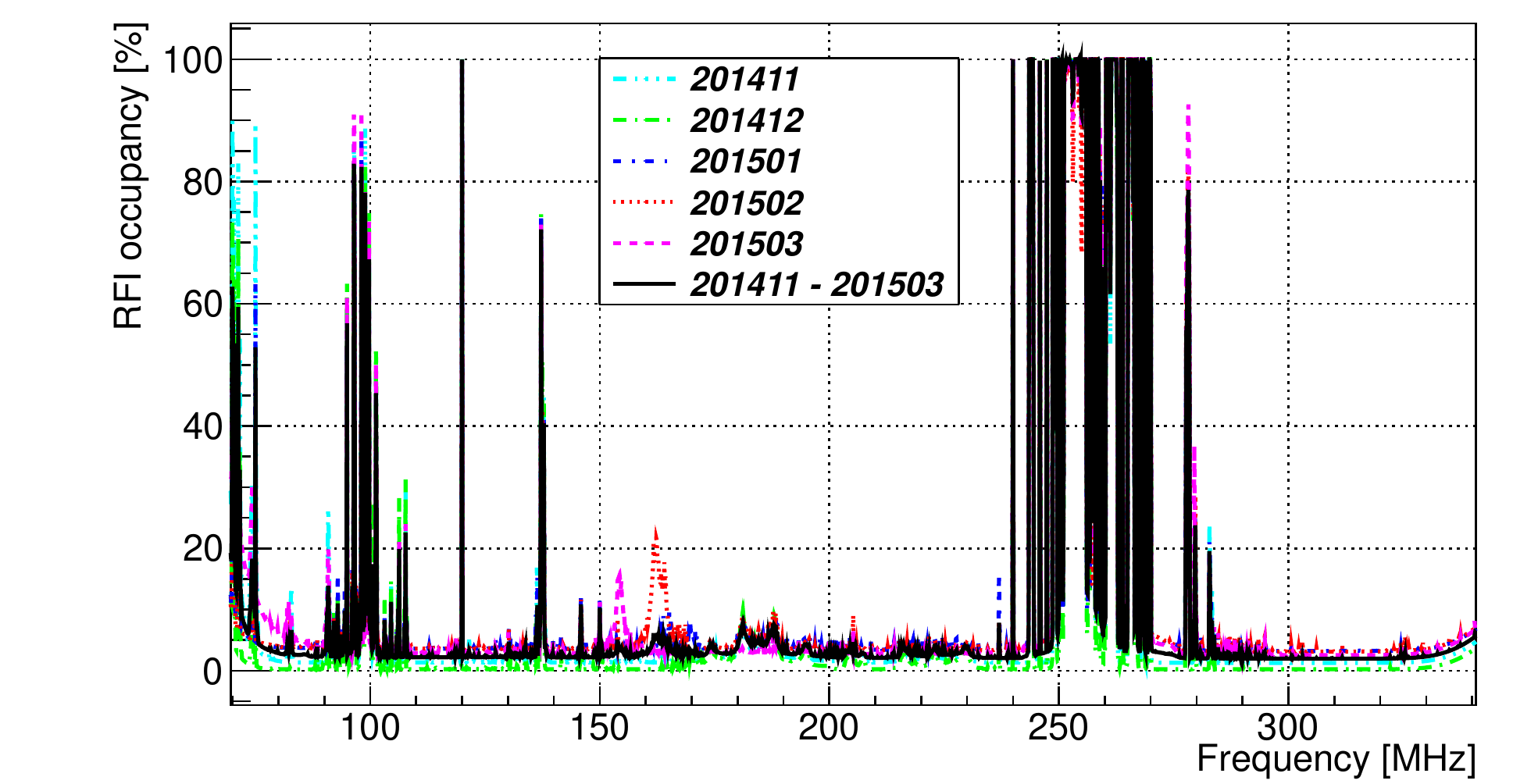}
\caption{Mean nighttime RFI occupancy (i.e. fraction of all data that was flagged) for each month of observations (dashed/dotted lines) and for the entire 201411 - 201503 dataset (black solid curve). The most significant excision rates can be seen in the military satellite band (242-272\,MHz) with 100\% occupancy in multiple channels and at Orbcomm satellite communication band (137-138\,MHz) at $\sim$70\% rate. Although the FM band (88-108\,MHz) looks significantly occupied, in fact only 3 channels exceed 80\% occupancy, 3 other exceed 40\% and 4 more exceed 20\%. The broadband peaks in the magenta (150-160\,MHz) and red curves between (160-170\,MHz) turned out to be an instrumental effect, which was fixed. The channel at 120\,MHz was entirely excised due to known internal RFI in the spectrometer.} 
\label{fig_mean_occupancy}
\end{figure}

Figure \ref{fig:example_spectra} shows example data captured by the system with typical nighttime data when the Galactic center is above and below the horizon.
Also shown in the figure is an example of an extreme RFI event, discussed further in Section \ref{sec:results}.

\subsection{RFI detection}

Data were recorded in approximately 50\,s blocks ($\approx$25\% of time was recorded on the calibrator source) and processed to detect RFI at the highest available time and frequency resolution.
We used the AOFlagger \cite{2012A&A...539A..95O}\footnote{\url{http://ascl.net/1010.017}} software as an automated system to detect RFI and other anomalies in the data.
The algorithm used by AOFlagger is sophisticated, but in essence it looks for outlier data points, or clusters of data points, in the time and/or frequency axis, as compared to the statistics of the entire data set.
In this case, each 50\,s block of data is processed independently and flags that were generated for each block of data were used in subsequent processing.

\section{Results and analysis}
\label{sec:results}
In the following analysis we use nighttime data unless otherwise specified because highly variable solar activity during the daytime can generate false positive detections of RFI.
Radio astronomy can of course be performed during the day, however strong radio emission from the sun can be problematic for some types of observations.
For this reason, the key (non-solar) science projects for the SKA with the most stringent sensitivity requirements are likely to be executed only at night.

Known sources of RFI at the MRO include\footnote{\url{http://www.acma.gov.au/theacma/australian-radiofrequency-spectrum-plan-spectrum-planning-acma}}:
persistent satellite-based RFI between 242 and 272\,MHz; intermittent relatively powerful transmissions from Orbcomm and other satellites around 137\,MHz; infrequent aircraft communications in the 118 to 137\,MHz band.
In addition, atmospheric conditions can sometimes be favorable for ``tropospheric ducting'' and other phenomena that allow very long-distance propagation to occur, hence for FM radio, digital television (DTV) and Digital Audio Broadcast (DAB, i.e. digital radio) from distant cities to reach the MRO.
Nearby lightning also produces broadband radio signals, which were observed and flagged during multiple summer afternoons and evenings, sometimes covering significant fractions of nights.
We analyze the frequency of such events in Section \ref{sec:extreme_events}.
Short bursts (timescales of seconds) of RFI are observed due to reflections of transmitted signals (e.g. FM radio) from aircraft or meteor trails (Fig.~\ref{fig_reflection_example}).
Finally, occasional powerful transmissions from satellites (e.g. Orbcomm) or aircraft can saturate the receiver and therefore entire band gets saturated and data are consequently excised.

\begin{figure*}[]
\centering
\includegraphics[width=\textwidth]{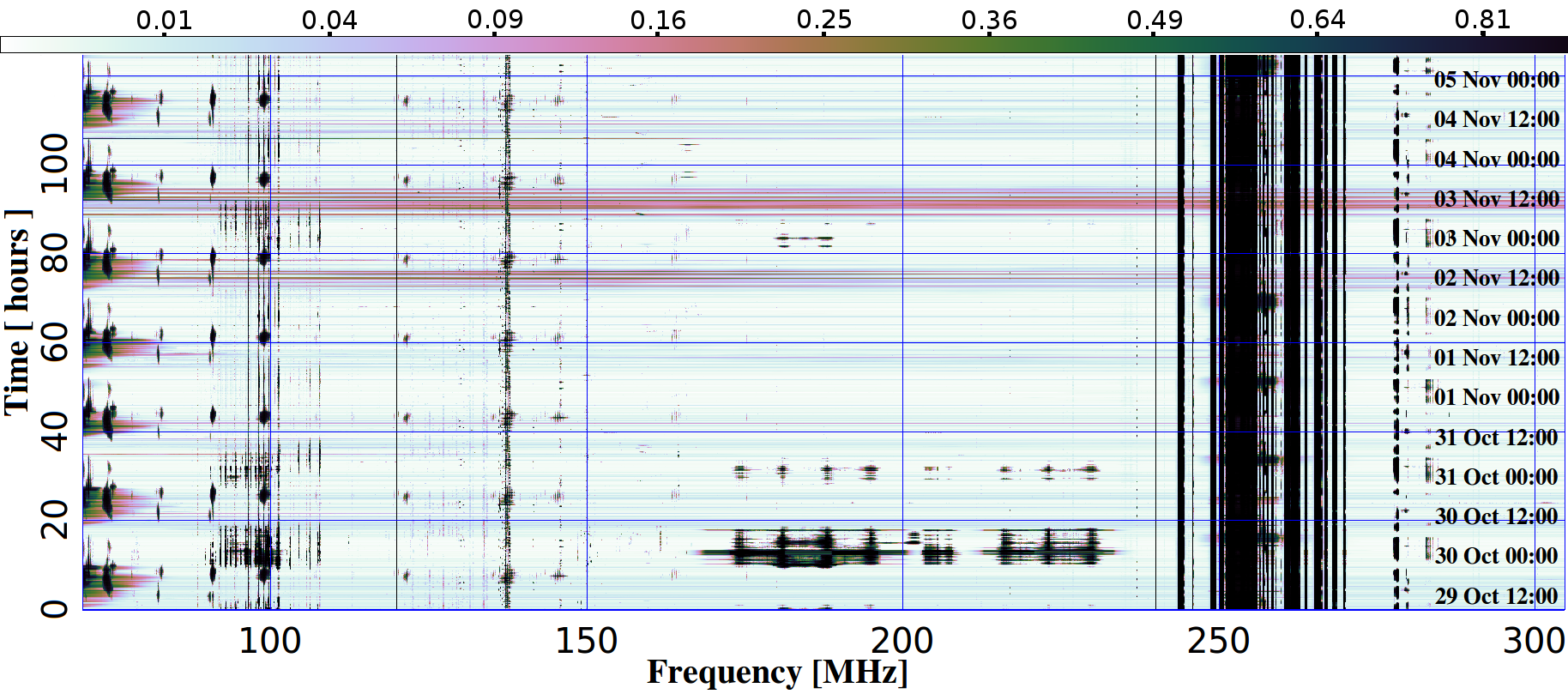}
\caption{Occupancy averaged in 10\,min bins as a function of frequency (horizontal axis) and time (vertical axis) over a period of 6\,days starting at 2014-10-29 08:00 (AWST). Besides the features already described in Figure~\ref{fig_mean_occupancy}, significant RFI excision of digital TV/radio and FM radio can be observed during the nights of strong (2014-10-29/30) and weak (2014-10-30/31 and 2014-11-02/03) tropospheric ducting events (see Tab.~\ref{tab_data_stat}). It can also be noted that FM radio signals become stronger around sunsets. The light magenta horizontal lines around 12:00 on 2014-11-02 and 2014-11-03 (occupancy of the order of a few up to 20 percent) across the entire band result from the significant solar activity on these two days, which was also flagged as RFI.}
\label{fig_occupancy_dyn_spec}
\end{figure*}

\subsection{RFI occupancy vs frequency}
Mean nighttime RFI occupancy for each month and for the entire analyzed period is shown in Figure \ref{fig_mean_occupancy}. Overall, occupancy in the ``clean channels'', which constitute most of the band, is below 5\%. 
Nevertheless, there are multiple channels which are practically completely excluded from any astronomical data analysis. The most significant excision rates can be seen in the military satellite band (242-272\,MHz) with 100\% rate in multiple channels and at Orbcomm satellite communication band (137-138\,MHz) at $\sim$70\% rate. 
Relatively high power in a single channel (e.g. Orbcomm satellite or aircraft communications) can sometimes saturate the receiver and cause all the spectra collected during the transmission to be excised. 
Moreover, similar but less powerful events can cause undesired distortions in the noise floor; it is therefore important to monitor the RFI environment also outside the observing band.
In Figure~\ref{fig_mean_occupancy} the FM band (88-108\,MHz) looks significantly occupied, but in fact only 3 channels (96.5\,MHz, 98.1\,MHz and 98.9\,MHz) exceed 80\% occupancy, 3 other (94.9\,MHz, 99.7\,MHz and 101.3\,MHz) exceed 40\% and 4 (90.9\,MHz, 90.9\,MHz, 100.7\,MHz, 106.3\,MHz) more exceed 20\%. 
All the channels exceeding 40\% occupancy were identified as 10 and 30\,kW transmitters on the list of  Geraldton transmitters at the matching frequencies. The channels exceeding 20\% occupancy match transmitters in the nearby ($\le$300\,km) towns of Cue, Yalgoo, Kalbarri, Mullewa or Meekatharra, but the association is much more uncertain as only some of them exceed 1\,kW power.
There are also channels in the DTV, DAB and FM bands which are significantly affected during conditions favoring long distance propagation via tropospheric ducting.
However, the mean excision rates in DTV and DAB at 174-230\,MHz (Fig.~\ref{fig_mean_occupancy}) are not significantly higher than the overall excision floor because the data in these channels were only occasionally excised during conditions favoring long-distance propagation (Fig.~\ref{fig_occupancy_dyn_spec} and Sec.~\ref{sec:extreme_events}).

Figure~\ref{fig_occupancy_dyn_spec} shows a representative 6\,day sample of the occupancy averaged in 10\,min intervals as a function of frequency and time.
This figure clearly shows that nights, besides occasional tropospheric ducting events (mostly in the summer time) discussed in Section~\ref{sec:extreme_events}, are typically very quiet. Some of the nights (not shown in this figure) were significantly affected by thunderstorms near the MRO, which caused significant (up to $\sim$80\%) excision rates across the entire band.
However, thunderstorms in this area occur almost exclusively in the summer. On the other hand, significant excision rates across the entire band in the daytime data were usually caused by solar activity (Fig.~\ref{fig_occupancy_dyn_spec}).

\begin{figure*}[!t]
\centering
\subfloat{\includegraphics[width=0.335\textwidth]{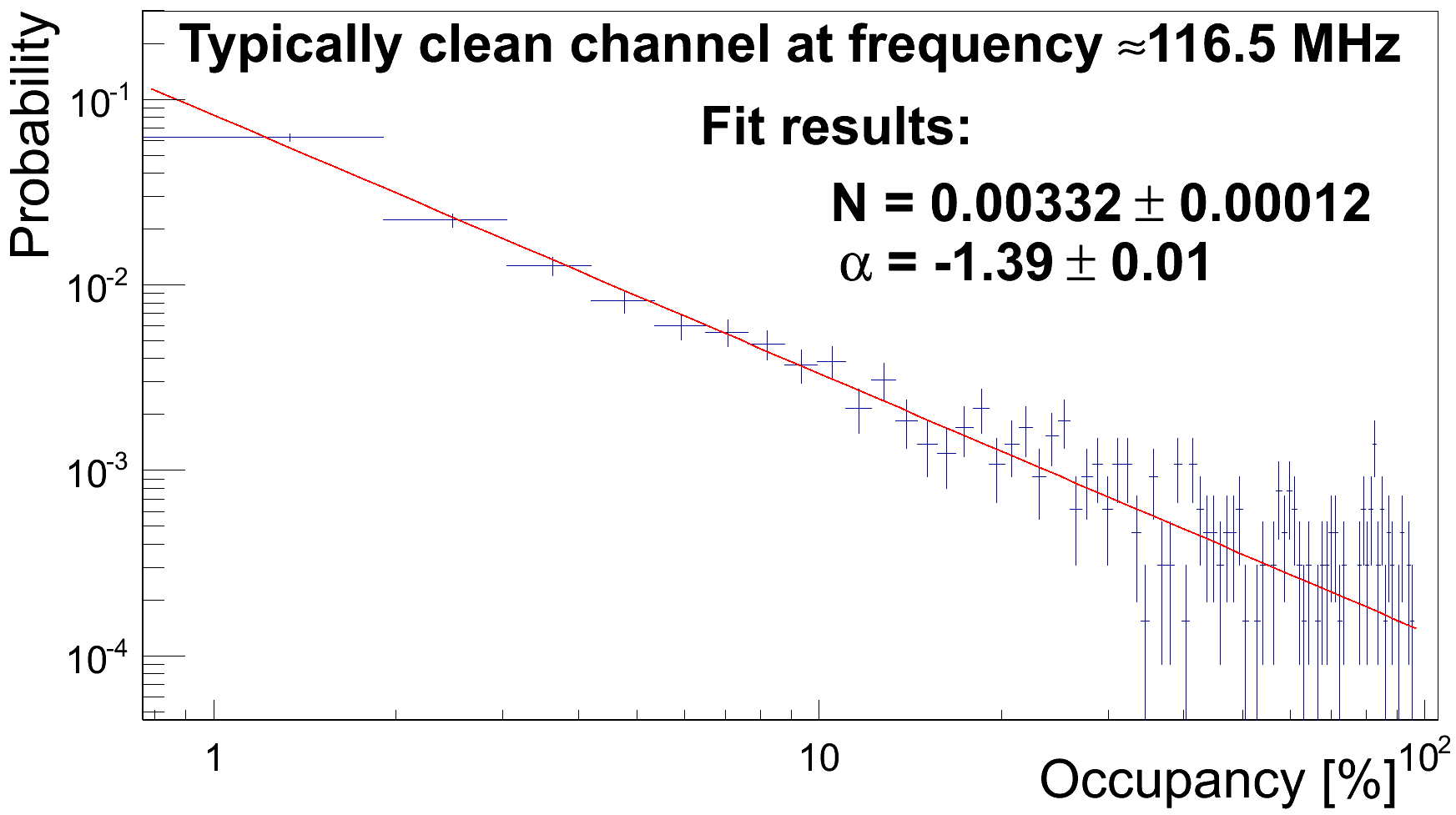}%
\label{fig_occupancy_distrib_CLEAN}}
\hfil
\subfloat{\includegraphics[width=0.32\textwidth]{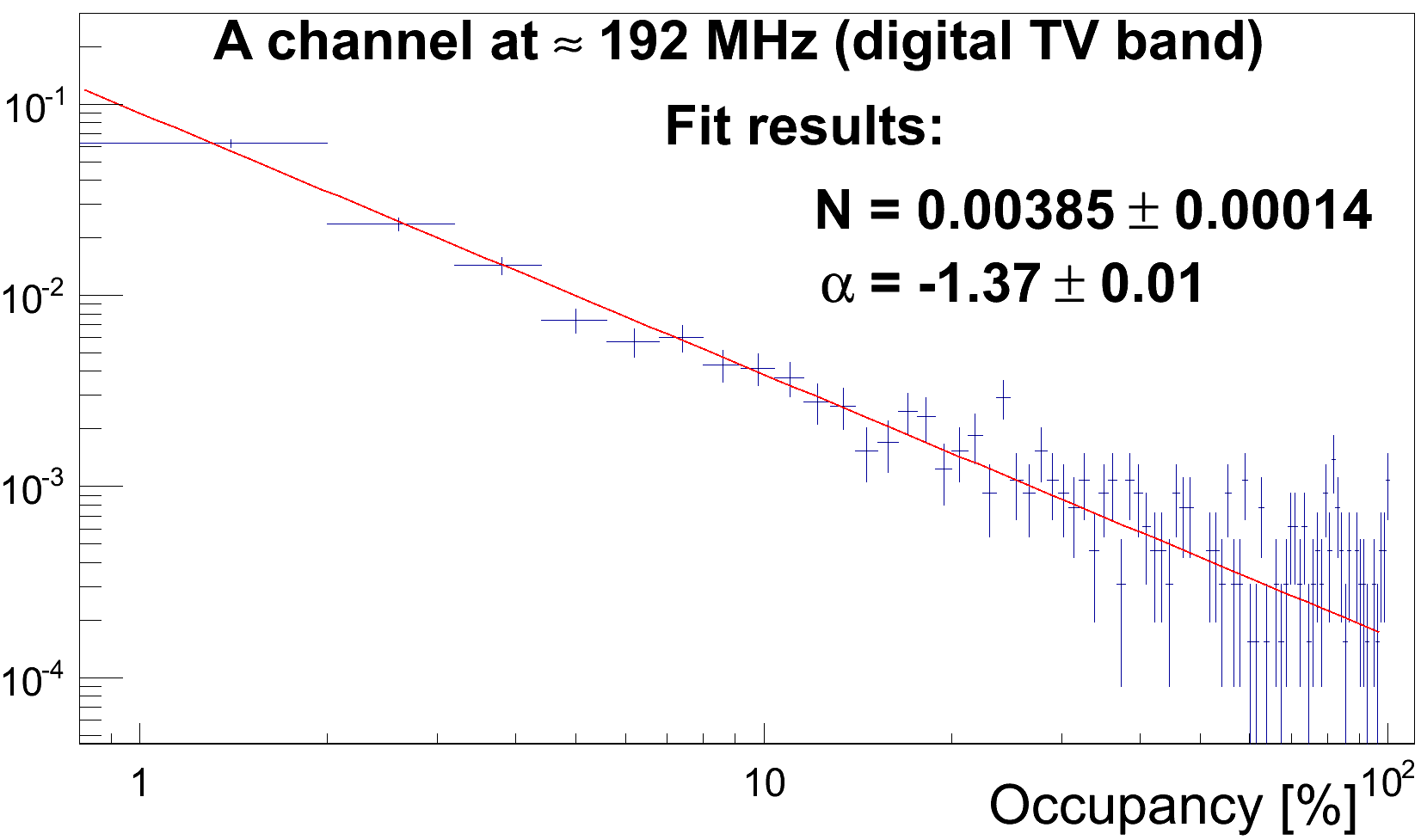}%
\label{fig_occupancy_distrib_DigTV}}
\hfil
\subfloat{\includegraphics[width=0.32\textwidth]{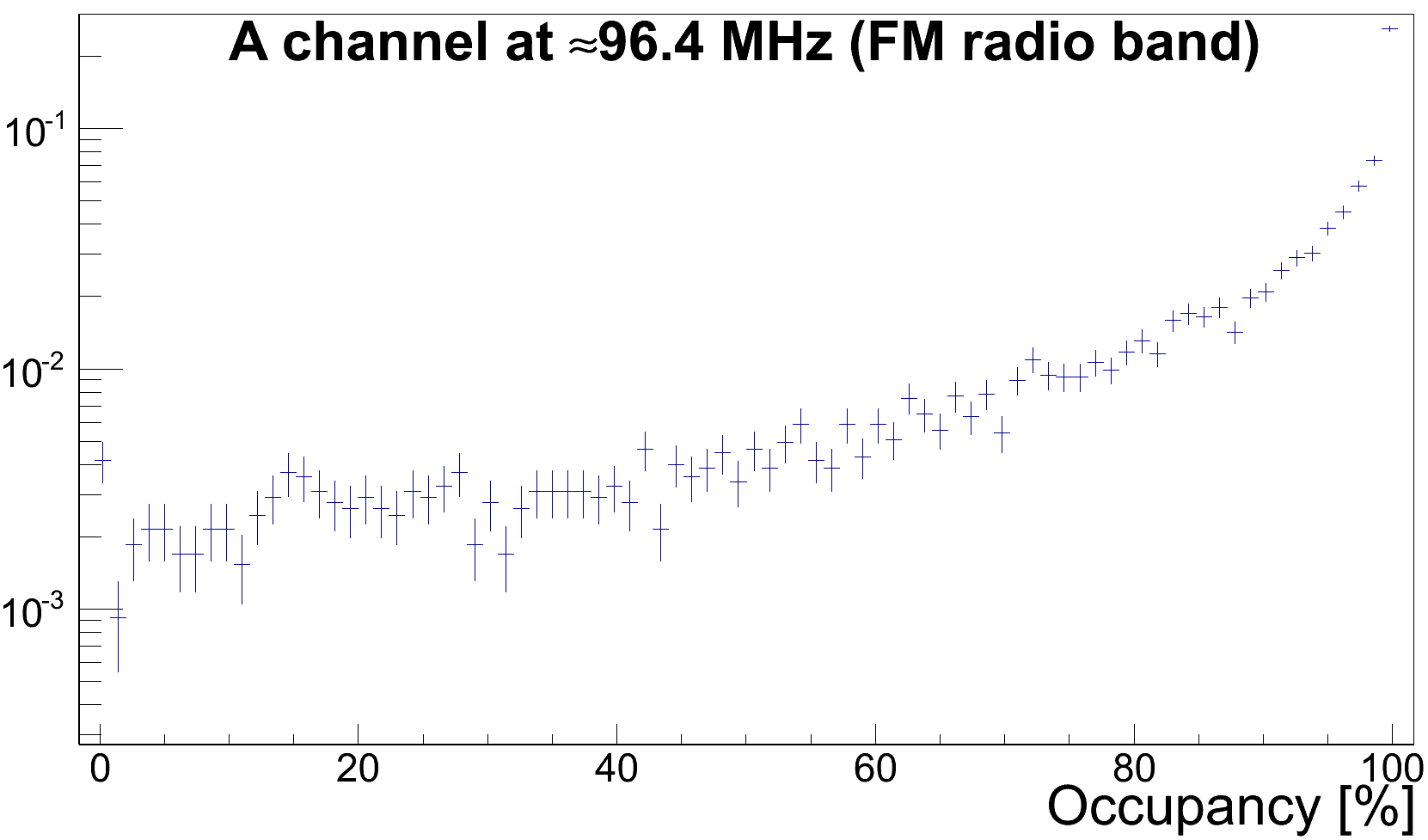}%
\label{fig_occupancy_distrib_FM}}
\caption{Occupancy distribution observed over the entire analyzed period in the three representative 117.2\,kHz channels. Left: a typically clean channel at 116.5\,MHz. Center: A channel at 192\,MHz which falls within the broadcast frequency of the DTV, but is also relatively clean because significant DTV power can only be observed during extreme events (Sec.~\ref{sec:extreme_events}). These two channeles were well parameterized with a power law function N$\times$(Occupancy[\%]/10\%)$^{\alpha}$. However, the distribution in a relatively occupied ($\approx$80\% on average) FM channel (96.4\,MHz) has a different shape with a significant chance ($\approx$20\%) of 100\% occupancy, which could not be fitted with a simple power law.}
\label{fig_occupancy_distrib}
\end{figure*}

\subsection{RFI occupancy statistics}
\label{sec:rfi_occ_stat}

\begin{figure}[]   
\centering
\includegraphics[width=\columnwidth]{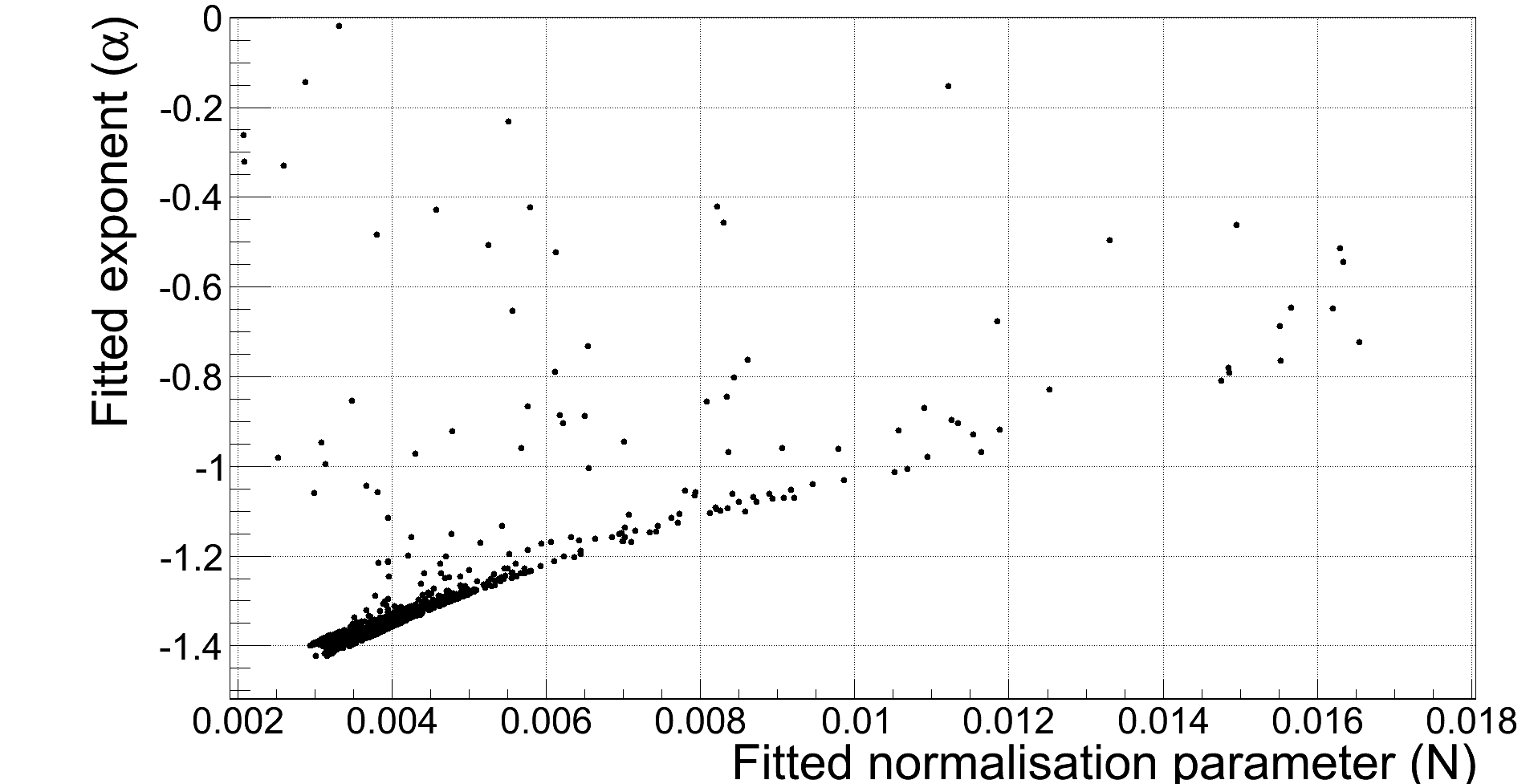}
\caption{Scatter plot of power law parameters characterizing occupancy distributions in all ``typically clean'' channels.}
\label{fig_fitted_parameters}
\end{figure}

The occupancy data (like in Figure~\ref{fig_occupancy_dyn_spec}) for the entire analyzed period enabled us to study distributions of occupancy in 10\,min intervals.
We have analyzed occupancy distributions for all the frequency channels in the analyzed band and the representative examples are shown in Figure~\ref{fig_occupancy_distrib}. 
The three channels were selected as: a ``clean channel'' at $\approx$116.5\,MHz devoid of significant RFI, an FM channel at $\approx$96.4\,MHz and a DTV channel at $\approx$192\,MHz.
Distributions in relatively clean frequency channels (including those which like DTV or DAB are occupied by RFI only during extreme tropospheric ducting events) were fitted with a power law ( N$\times$(Occupancy[\%]/10\%)$^{\alpha}$ ).
However, occupancy distributions for highly occupied frequency channels (e.g. 96.4\,MHz in the FM band) cannot be fitted with a simple power law (Fig.~\ref{fig_occupancy_distrib_FM}). 
The scatter plot of fitted parameters of the power law fits to all ``typically clean'' frequency channels is shown in Figure~\ref{fig_fitted_parameters}.
The vast majority of points in the Figure~\ref{fig_fitted_parameters} are clustered around $N=0.004$, confirming that the typical RFI occupancy is much less than 10\%.
The parameters fitted to ``non-clean'' channels lie outside the ranges of the figure, but such highly ``polluted'' channels ($>$40\%) can be considered as lost to most astronomical observations.
Such a procedure enables us to make estimates of amount of data lost due to RFI in every single frequency channel, which is a very important input to the design and observing strategies of the SKA-Low telescope to be built in the same location.


\subsection{RFI power statistics}
\label{sec:rfi_power_stat}

\begin{figure*}[!t]
\centering
\subfloat{\includegraphics[width=0.335\textwidth]{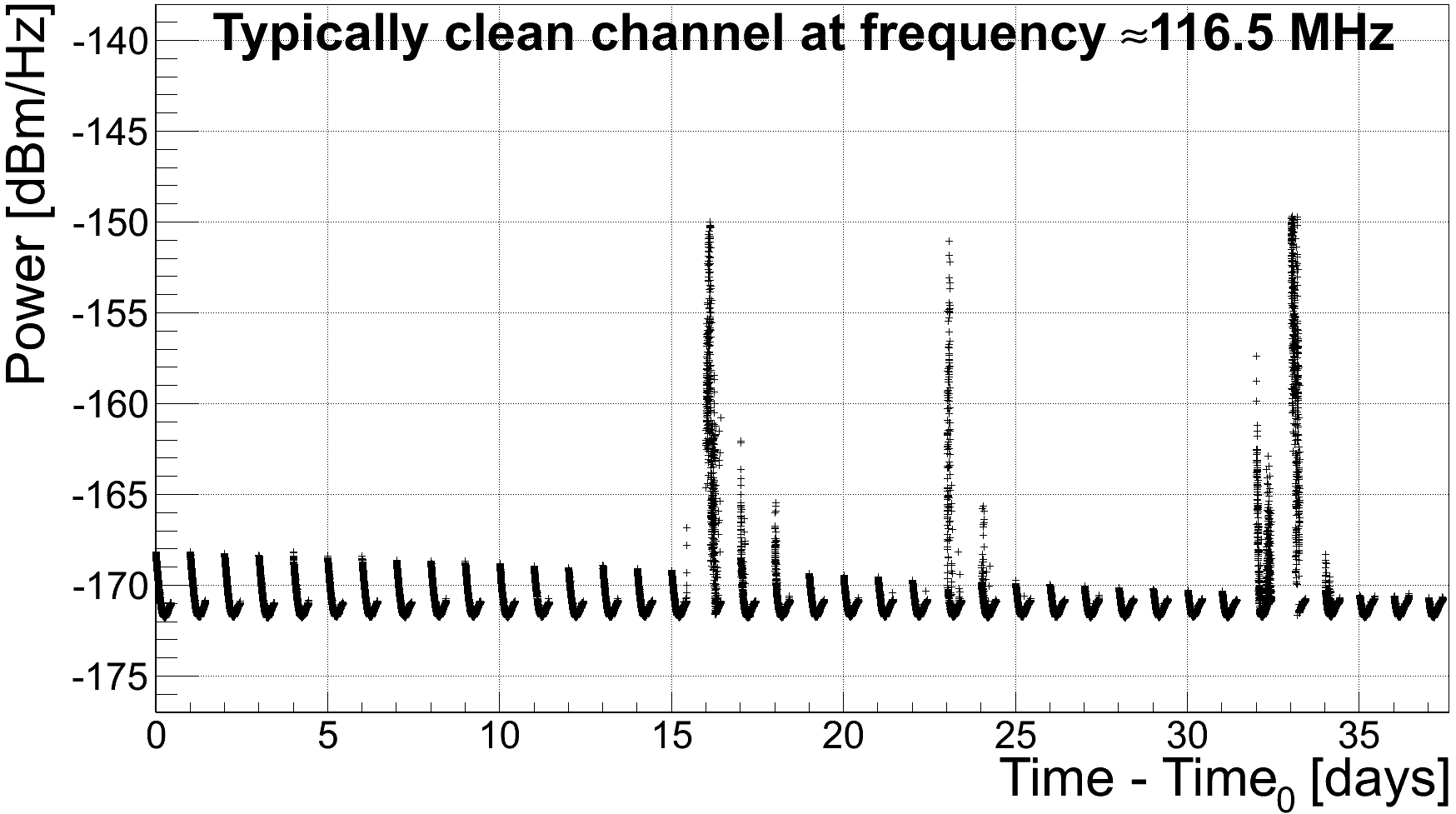}%
\label{fig_clean_chan_vs_time}}
\hfil
\subfloat{\includegraphics[width=0.32\textwidth]{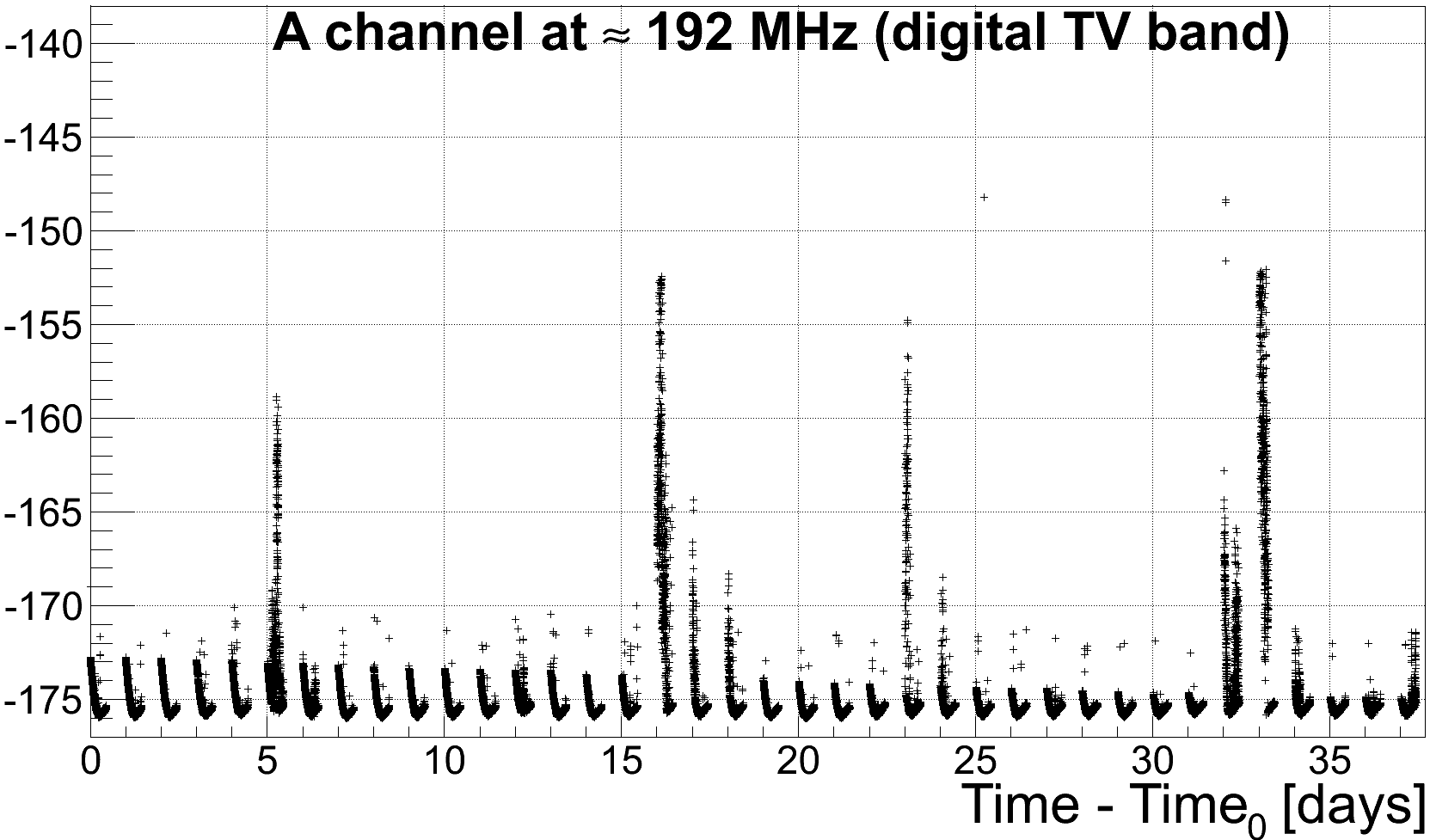}%
\label{fig_DTV_vs_time}}
\hfil
\subfloat{\includegraphics[width=0.32\textwidth]{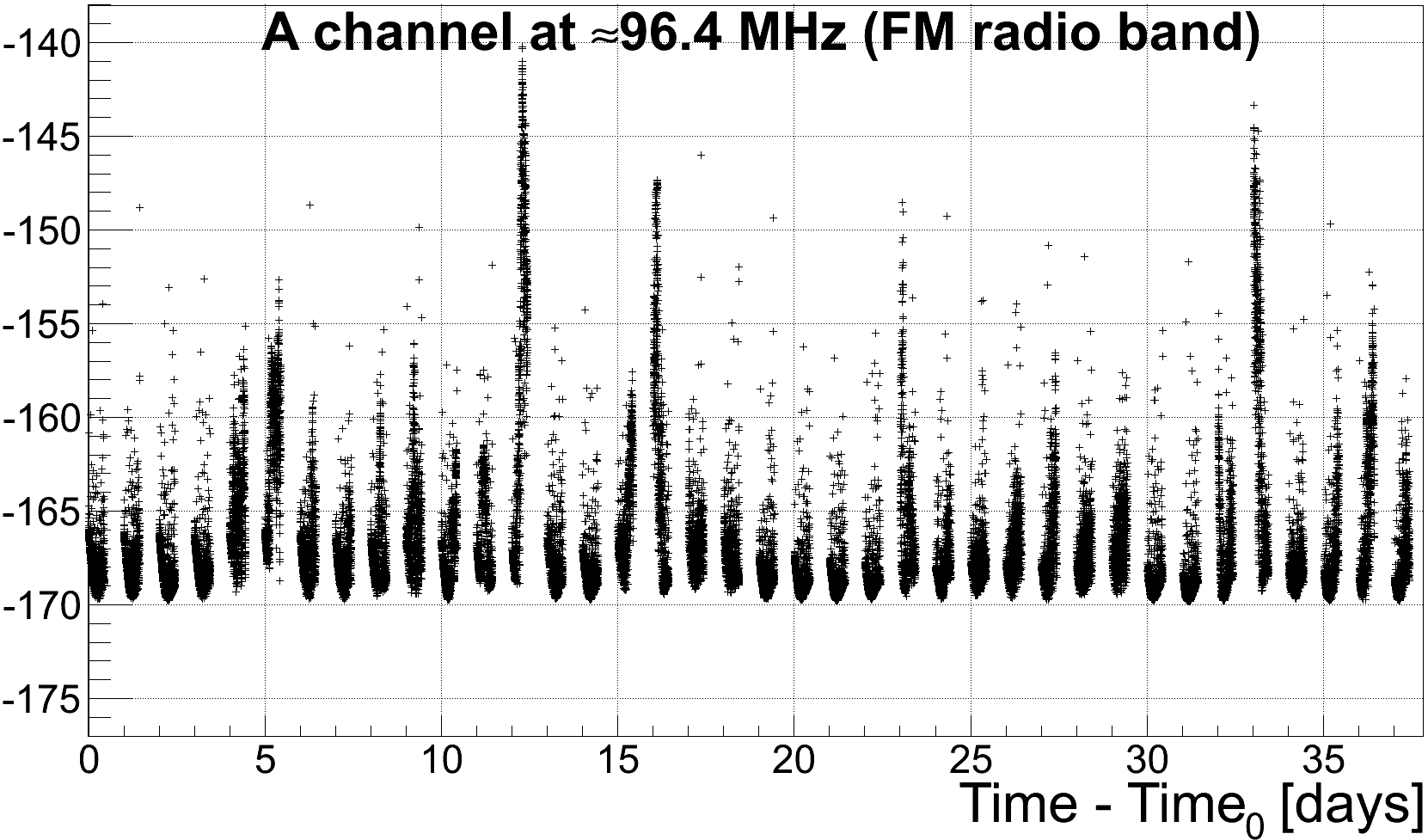}%
\label{fig_FM_vs_time}}
\caption{Examples of the power measured in a single 117.2\,kHz channel over 38 days of data collected in October and November 2014. Left: a typically clean channel at 116.5\,MHz. Center: A channel at 192\,MHz which falls within the broadcast frequency of digital TV. Right: A channel at 96.4\,MHz, which is within the FM radio band.}
\label{fig_sim}
\end{figure*}

In order to study distribution of observed RFI power, the raw spectra recorded by the spectrometer were reduced to a more manageable volume by calculating maximum spectrum within 35\,s of antenna data. 
In the next step, the spectra were calibrated in temperature units according to our calibration procedure \citep{2015PASA...32....4S,2015ApJ...813...18S}, which were subsequently converted to dBm/Hz based on the frequency bin width of 117.2\,kHz.
The calibrated power as a function of time in the same three representative 117.2\,kHz frequency channels (Sec.~\ref{sec:rfi_occ_stat}) is shown in Figure~\ref{fig_sim}. 
The ``clean channel'' (116.5\,MHz) shows bursts of significant RFI during several nights, which were verified to originate from the thunderstorms near the MRO. 
The DTV channel (192\,MHz) also shows the same bursts as the ``clean channel'', but it also shows at least two other significant power increases during the 6th and the last night, which we attribute to tropospheric ducting events (see Fig.~\ref{fig:example_spectra} for example spectra from the 6th night).
Finally, the FM channel (96.4\,MHz) shows power bursts during the same nights as the DTV channel, but it also has many occasional high power points over all nights.
The noise floor of the FM channel is higher than the other two channels due to much larger Galactic noise at this frequency (the brightness temperature of Galactic emission scales approximately as $ \nu^{-2.6}$).

\begin{figure*}[!t]
\centering
\subfloat{\includegraphics[width=0.335\textwidth]{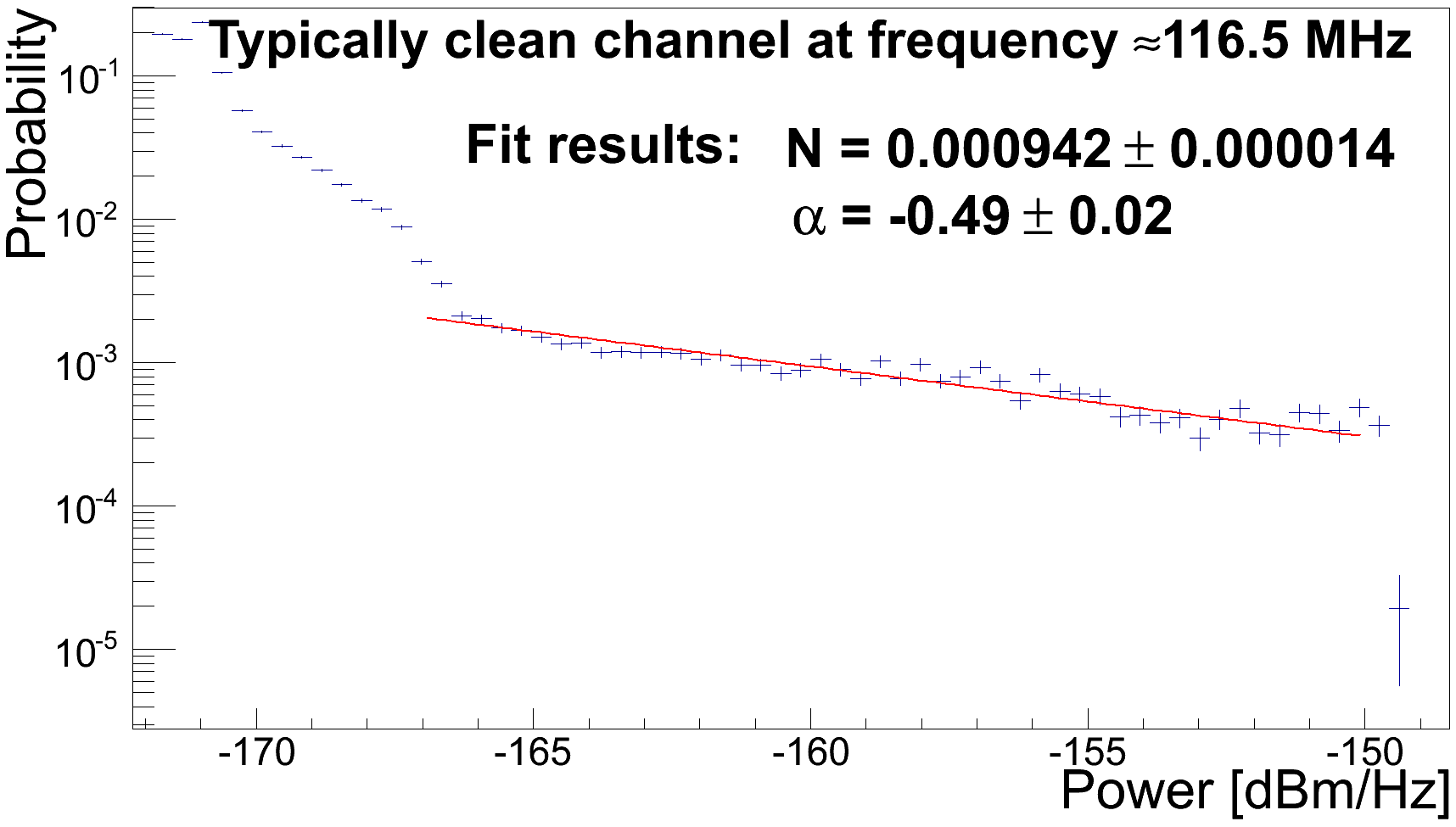}%
\label{fig_power_distrib_CLEAN}}
\hfil
\subfloat{\includegraphics[width=0.32\textwidth]{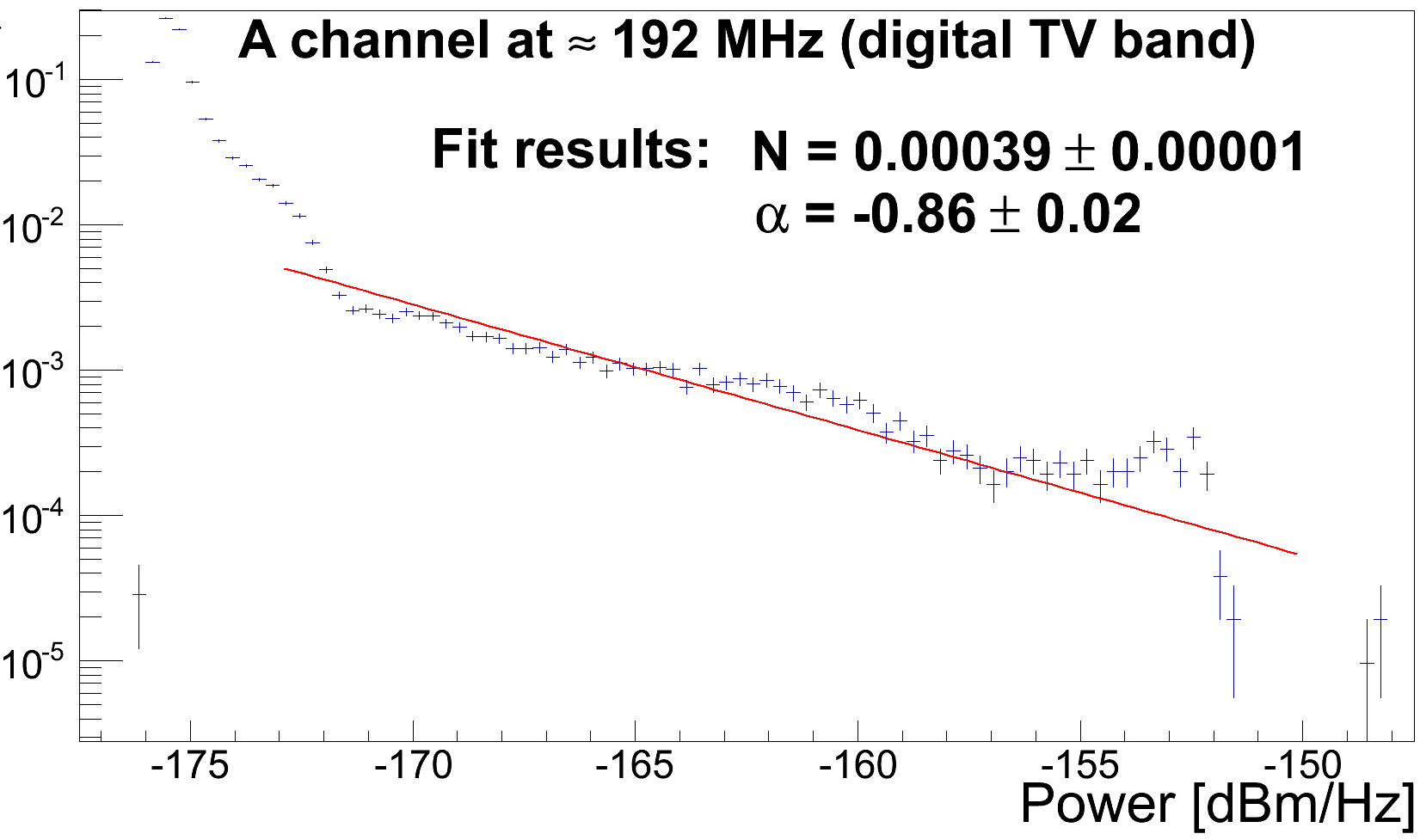}%
\label{fig_power_distrib_DigTV}}
\hfil
\subfloat{\includegraphics[width=0.32\textwidth]{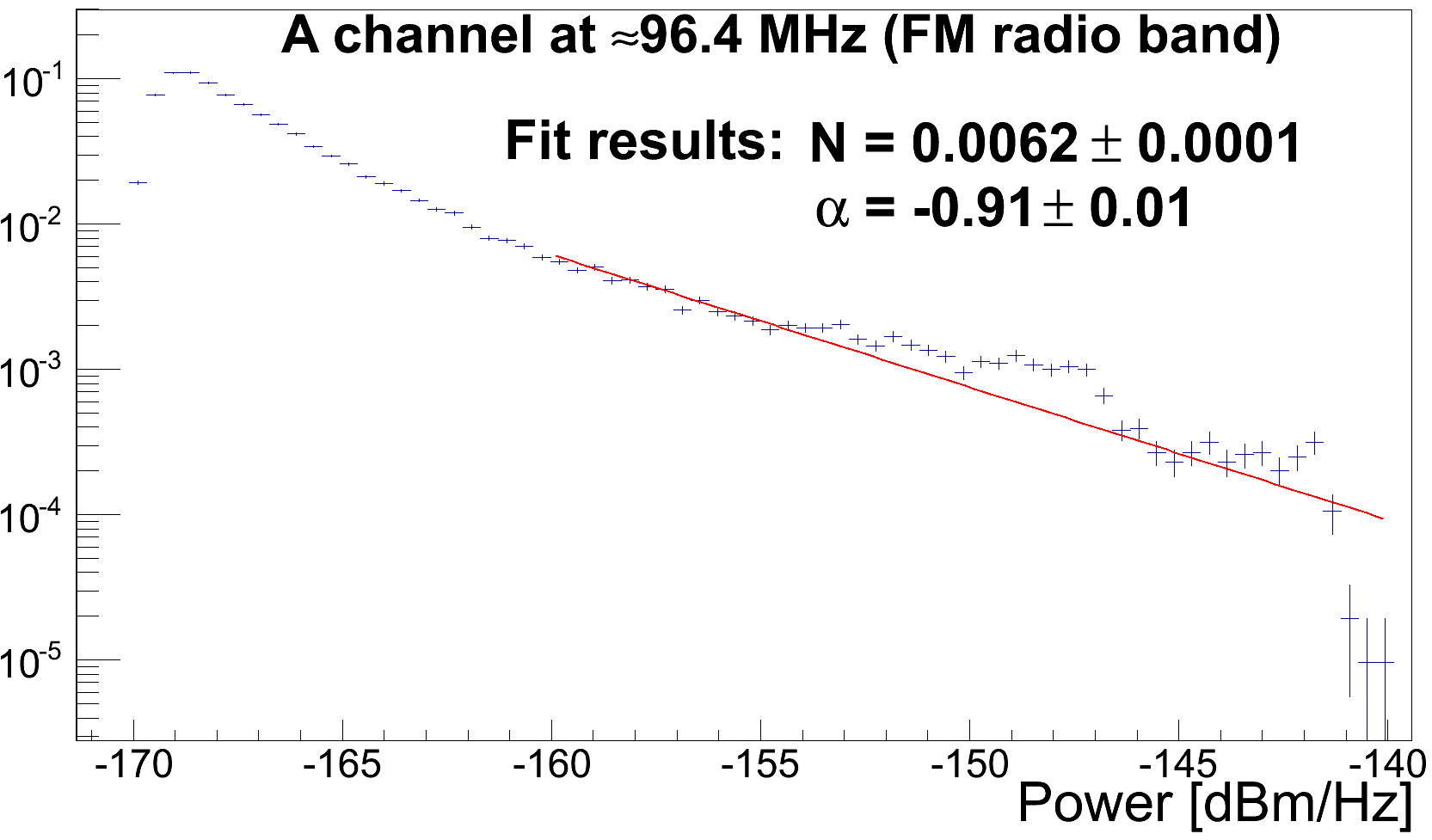}%
\label{fig_power_distrib_FM}}
\caption{The normalized distribution of power observed over the entire analyzed period in the same 117.2\,kHz channels as listed in Fig~\ref{fig_sim} with fitted power law function N$\times$(Power[mW/Hz]/$10^{-16}$[mW/Hz])$^{\alpha}$. The lower limit of the fit range was chosen according to maximum power of the Galactic noise at a given frequency. We have identified that the high power tail (e.g. around -140\,dBm/Hz in the right figure) is due to reflections of signals from remote transmitters of aircraft or meteorites.}
\label{fig_power_distrib}
\end{figure*}

The normalized distributions of power observed in these three channels over the entire observation period are presented in Figure~\ref{fig_power_distrib}.
The structures of these distributions are very similar. At low power the distributions are dominated by Galactic noise, which cuts-off at about -164\,dBm/Hz in FM channel and -172\,dBm/Hz in the DTV channel. We fitted the higher power parts of the power distributions with a power law function and results of the fits are also shown in Figure~\ref{fig_power_distrib}.
Interestingly, the exponent fitted to the power distribution in the ``clean channel'' is almost exactly 0.5, as expected for the free-propagating radio signals from uniformly distributed lightning of constant power occurring at random distances from the antenna. 
In the cases of FM and DTV channels the fitted exponents are closer to one, which, as expected, corresponds to larger suppression than $\sim1/r^2$ free-space propagation.
We have applied this approach to all the frequency channels in 70-300\,MHz band in order fully characterize RFI power probability distribution and can provide the data upon request. 

\subsection{Extreme events}
\label{sec:extreme_events}
In Figure \ref{fig:example_spectra} an example is shown of an extreme RFI event that includes clear signal around 205\,MHz as well as clear DTV and FM radio signals (Western Australia switched entirely to digital TV in April 2013).
The two 2\,MHz wide signals at 205\,MHz are DAB signals. At the time the data were taken, the only DAB broadcast site in Western Australia was from Perth, about 600\,km away from the MRO (the line-of-sight coverage of Perth transmitters is of the order of 80\,km).
Based on the appearance of DTV/DAB and FM radio signals in the spectra, we have counted number of such extreme RFI events and classified them according to their relative power above the typical Galactic noise level at corresponding frequencies (Tab.~\ref{tab_data_stat}).
The number of such events was higher during the hottest months (October, November, December and January 2015) and declined in February and March 2015, which is consistent with expectations for tropospheric ducting \cite{tropo_ducting}.

\begin{table}
\begin{center}
\caption{Statistics of analyzed nights with the number of thunderstorms (usually in the evenings, but sometimes until 2\,am) and observed extreme propagation events, which were classified according to their power above the Galactic noise background at a given frequency in DTV/DAB bands.}
\begin{tabular}{@{}ccccccc@{}}
\hline
Month & Nights & \begin{tabular}{@{}c@{}} Evening \\
storms \end{tabular} & \begin{tabular}{@{}c@{}} Strong \\
($>10$\,dB) \end{tabular} & \begin{tabular}{@{}c@{}} Moderate \\
($>3$\,dB) \end{tabular} & \begin{tabular}{@{}c@{}} Weak \\ ($<3$\,dB) \end{tabular} \\
\hline
2014-(10, 11) & 38 & 7 & 2 & 3 & 7 \\
2014-12 & 31 & 2 & 3 & 1 & 7\\
2015-01 & 31 & 11 & 2 & 1 & 3 \\
2015-02 & 28 & 13 & 0 & 3 & 3 \\
2015-03 & 16 & 5 & 0 & 0 & 2 \\
\hline\hline
\newline
\end{tabular}
\end{center}
\label{tab_data_stat}
\end{table}

\section{Conclusions}
We have analyzed approximately 5 months of low frequency radio data, collected between 2014 November and 2015 March inclusive from a single antenna spectrometer system installed at the MRO, to gather statistics on the magnitude and frequency of RFI.
Our data show that, with the exception of known satellite-based interferers, the site is extremely radio quiet even in the frequency bands containing television and radio broadcasts.
We have found that occupancy of several channels within the FM band exceeds 20\% due to RFI from relatively nearby (within 300\,km) high power transmitters (mainly in Geraldton).
However, most of the FM band is clean and its occupancy typically does not exceed 5\%. 

We have found that the major contributions to RFI in the summer are lightning from nearby thunderstorms and from unusual atmospheric conditions (tropospheric ducting) that allows television and radio signals to reach the site with relatively high received power.
We used digital television and digital radio bands to identify tropospheric ducting events.
The tropospheric ducting phenomenon occurs mostly during the hottest months.
Strong ducting events ($>$10\,dB above background sky signal) were identified only in 5\% of the nights; RFI from digital television and digital radio ($>$1\,dB above background sky signal) was detectable in 27\% of our data.
Within the studied frequency band (70-300\,MHz), atmospheric ducting mostly affects frequencies in broadcast radio and television bands whilst the lightning causes broadband radio emission that affects the entire band.
The thunderstorms are frequent in the area only during the summer, but they typically end at evening/early night and only in several cases they affected significant fractions of nighttime data.

Using calibrated data we studied the probability distributions of RFI power in several representative frequency channels.
The probability density function of RFI power can be fitted with a power law with exponents in the range -0.5 to 1.0.
This analysis extended to the entire band enables engineers and future observers to calculate the probability of strong RFI, including  that which might saturate a receiver.
We have also studied the occupancy distributions, which, at least for typically clean channels, can also be parameterized with a power law (a consequence of RFI power distribution). Such a study enables us to predict fraction of observing data lost due to RFI. 

Our analysis was performed on data collected from a single conical log-sprial antenna, which has a response that is suppressed near the horizon (gain within 5$\degree$ from the horizon is $\le$~-10~dBi). 
However it is still relatively sensitive compared to the far sidelobe response of radio telescope primary beams, which can be further suppressed by some 20-40\,dB. 
The remaining RFI signals which need to be considered are those coming from the sky such as communication satellites (e.g. Orbcomm), DTV/DAB and FM band reflections of aircraft or meteorites and ducted broadcast transmissions from remote transmitters during extreme tropospheric ducting events (seasonal).
Particularly, Orbcomm satellites with their relatively powerful intermittent transmissions can cause receiver saturation.

\section*{Acknowledgment}
We thank CSIRO and the MWA operations team for their ongoing support of BIGHORNS and its supporting infrastructure at the Murchison Radio-astronomy Observatory. We acknowledge the Wajarri Yamatji people as the traditional owners of the Observatory site.
This research was conducted by the Australian Research Council Centre of Excellence for All-sky Astrophysics (CAASTRO), through project number CE110001020.
The International Centre for Radio Astronomy Research (ICRAR) is a Joint Venture between Curtin University and the University of Western Australia, funded by the State Government of Western Australia and the Joint Venture partners.

\newcommand{\pasa}{PASA}
\newcommand{\apj}{ApJ}
\newcommand{\aj}{AJ}
\newcommand{\apjl}{ApJ}
\newcommand{\aap}{A\&A}
\newcommand{\mnras}{MNRAS}
\newcommand{\aaps}{A\&AS}
\newcommand{\pasp}{PASP}
\newcommand{\physrep}{PhR}
\newcommand{\araa}{ARA\&A}
\bibliographystyle{IEEEtran}
\bibliography{IEEEabrv,refs}

\end{document}